\documentclass[12pt,draftclsnofoot, onecolumn]{IEEEtran}
\usepackage{graphicx}
\usepackage{color}
\usepackage{url}
\usepackage{caption}
\usepackage{subcaption}
\begin{document}
\title{6G: The Next Frontier}
\author{Emilio Calvanese Strinati,~\IEEEmembership{Member,~IEEE}, Sergio Barbarossa,~\IEEEmembership{Fellow,~IEEE}, \\  Jos\'e Luis Gonzalez-Jimenez,~\IEEEmembership{Member,~IEEE}, Dimitri Kt\'enas,~\IEEEmembership{Member,~IEEE}, Nicolas Cassiau,~\IEEEmembership{Member,~IEEE}, Luc Maret,~\IEEEmembership{Member,~IEEE}, \\ C\'edric Dehos,~\IEEEmembership{Member,~IEEE}\\
\thanks{Emilio Calvanese Strinati, Jos\'e Luis Gonzalez-Jimenez, Dimitri Kt\'enas, Nicolas Cassiau, Luc Maret and C\'edric Dehos are with CEA-LETI; Sergio Barbarossa is with the Department of Information Engineering, Electronics, and Telecommunications,
Sapienza University of Rome, Via Eudossiana 18, 00184,
Rome, Italy. E-mail: sergio.barbarossa@uniroma1.it}}

\maketitle
\begin{abstract}
5G  networks  represent  a  breakthrough  in  the  design  of  communication  networks,  for  their  ability to provide a single platform enabling a variety of different services, from enhanced mobile broadband communications  to  virtual  reality,  automated  driving,  Internet-of-Things,  etc.  Nevertheless,  looking  at the  increasing  requests  for  new  services  and  predicting  the  development  of  new  technologies  within a  decade  from  now,  it  is  already  possible  to  envision  the  need  to  move  beyond  5G  and  design  a new  architecture  incorporating  new  technologies  to  satisfy  new  needs  at  both  individual  and  societal level.  The  goal  of  this  paper  is  to  motivate  the  need  to  move  to  a  sixth  generation  (6G)  of  mobile communication  networks,  starting  from  a  gap  analysis  of  5G,  and  predicting  a  new  synthesis  of  near-future services, like holographic communications, high precision manufacturing, a pervasive introduction of artificial intelligence and the incorporation of new technologies, like sub-THz or Visible Light Communications (VLC), in a truly 3-dimensional (3D) coverage framework, incorporating terrestrial and aerial radio access points to bring cloud functionalities where and when needed on demand. 
\end{abstract}
%
%
\section{Introduction} 


\textcolor{black}{In 1926, the visionary Nikola Tesla stated: ``When wireless is perfectly applied, the whole Earth will be converted into a huge brain ...''. In 2030, pushed by fundamental needs at the individual as well as societal level, and based on the expected advancements of Information and Communication Technologies (ICT), Tesla's  prophecy may come to reality and 6G will play a significant role in this advancement by providing an ICT infrastructure that will enable the end users to perceive themselves as surrounded by a ``huge {\it artificial} brain'' providing virtual zero latency services, unlimited storage, and immense cognition capabilities.} 

5G networks already represent a significant leap forward in the realization of this grand vision. With respect to previous generations, rather than just improving the communications capabilities, 5G provides a communication infrastructure enabling a variety of services, or {\it verticals}, from Enhanced Mobile Broadband communication to Industry 4.0, automated driving, massive machine type communications, etc. In 2018, intensive successful testing, proof-of-concepts and trials \cite{5GCHAMPION2018}  have supported the launch in 2019 of the  fifth generation (5G) services, which will fundamentally transform current industries, create new industries, and impact societies and revolutionize the way people connect with everything and everything connects to people and things.
The services enabled by 5G networks are characterized by very diverse sets of key performance indicators (KPI's), so that the design of a single platform enabling all of them in an efficient manner is a very challenging task. The approach taken by 5G to address this challenge builds on the exploitation of {\it softwarization} and {\it virtualization} of network functionalities.
In parallel, the accommodation of stringent requirements in terms of data rate and latency, has required the introduction of millimeter-wave (mmW) communications, the exploitation of massive multiple-input/multiple-output (MIMO) links,  and the (ultra) dense deployment of radio access points. 


With its distinguishing connotation of being an enabler of very different services, 5G represents a major breakthrough with respect to previous generations. Given this context, the fundamental question we wish to address is: Given the enormous potentials of 5G networks and their foreseeable evolution, {\it is there a real motivation for thinking of 6G networks ?} If yes, \textit{what should there be in 6G that is not in 5G or in its long term evolution?} 
The \textit{``6G or not 6G''} debate has indeed already started. Academic, industrial and  research communities are working on the definition and identification of relevant key enabling technologies that might define the so called 'beyond 5G' (B5G) or sixth generation (6G) \cite{Netword2020-2015}, \cite{IEEEVTM-David2018}, \cite{Li}.  





\begin{figure}[ht]
\centering
\includegraphics[width=\columnwidth]{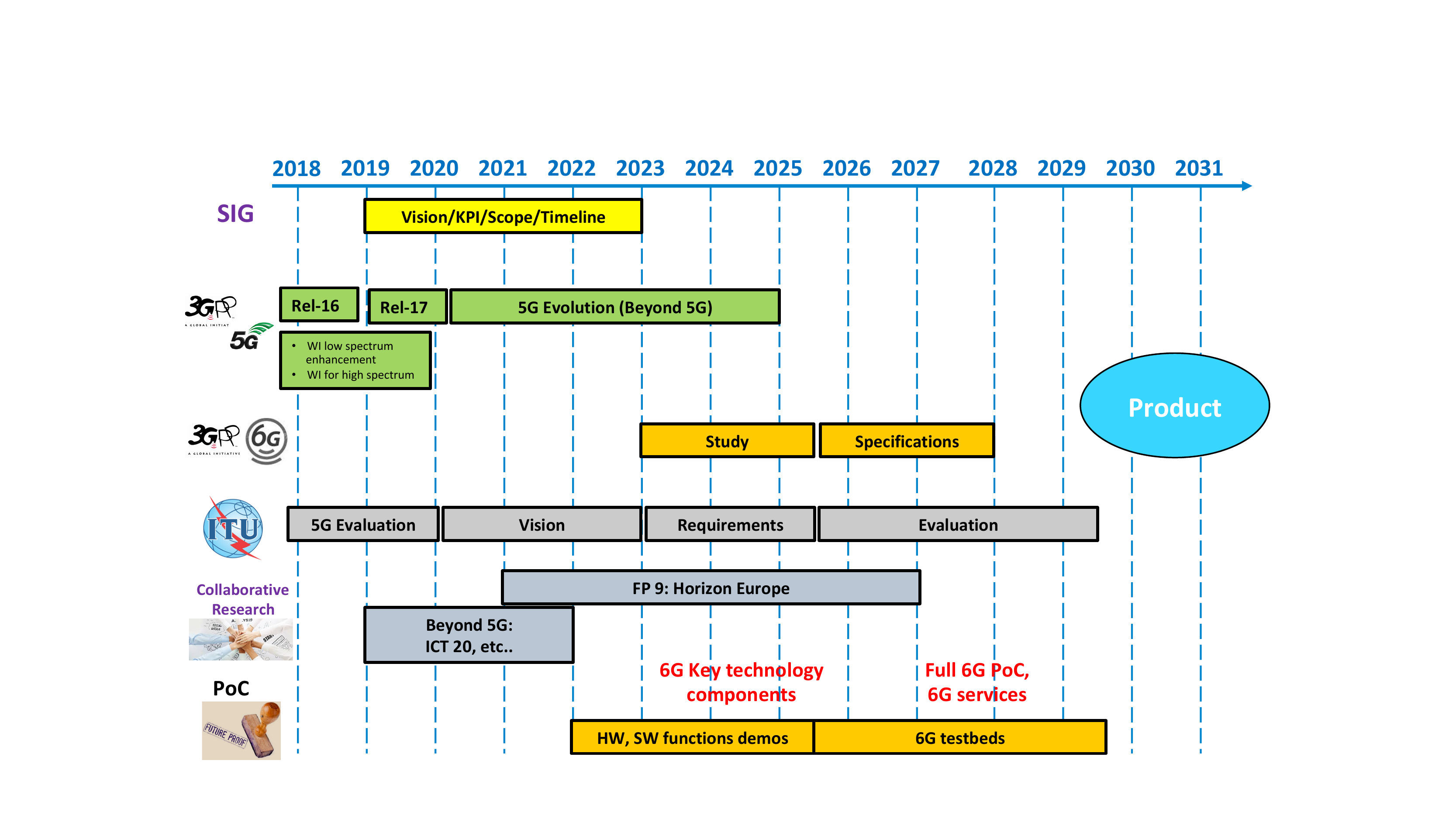}
\caption{Proposed 6G Roadmap.}
\label{fig:6GRoadmap} 
\end{figure}

In our vision a tentative 6G  roadmap is reported in Fig. \ref{fig:6GRoadmap}.
In this paper, we point out what we think will be the key distinguishing factors of 6G, starting, in Section II, with new services that motivate a shift towards 6G networks, and then identifying the major enablers of these new services. Section III focuses on a comprehensive vision of 6G, where a pervasive introduction of Artificial Intelligence and a holistic management of C4 resources are shown to represent a paradigm shift with respect to 5G. Section IV is then devoted to sub-THz and visible light communications. Finally, Section V draws some conclusions and highlights further developments.

\section{A new communication infrastructure to enable new services}
Typically, a new generation arises at the confluence of two major paths: a technological path that brings to maturity new groundbreaking technologies and a societal path that motivates the introduction of new services that cannot be efficiently offered by the current technologies. We start presenting new services and then we highlight some of their major enablers.

\subsection{New Services}
We envision the following new services that cannot be efficiently provided by the current development of 5G networks:

\textcolor{black}{
{\bf Holographic communications:} In ten years from now, the current ways of remote interaction between human beings will become obsolete, as new forms of interaction will arise leading to a true immersion into a distant environment. Five dimensions (5D) communications and services, integrating all human sense information (sight, hearing, touch, smell and taste) are expected to arise, together with holographic communications, thus providing a truly immersive experience \cite{IEEEVTM-David2018}. Holographic communications, employing multiple view cameras, will demand data rates in the order of Tbps \cite{Li}, which are not supported by 5G.} 

\textcolor{black}{
{\bf High-precision manufacturing:} The key objective in Industry 4.0 is to reduce the need for human intervention in industrial processes by using automatic control systems and communication technologies. 
In numerical terms, when applied to high-precision manufacturing, this translates to very high reliability - up to the order of $10^{-9}$ - and extremely low latency, in the order of $0.1$ to $1$ millisecond (ms) round trip time \cite{Berardinelli2018}. Furthermore, industrial control networks require real-time data transfer and strong determinism, which translates into a very low delay jitter, in the order of $1\, \mu$sec \cite{Berardinelli2018}.\\
}
\textcolor{black}{
{\bf Sustainable development and smart environments: } ICT technologies, incorporating wireless communications, cloud computing, and the Internet of Things (IoT), are expected to play a key role to drive global sustainability and improve quality of life. ICT can strongly contribute to improve health care, enable the development of smart cities, including the design of intelligent transportation and energy distribution systems. Achieving some of  these goals requires a pervasive sensing and a distributed decision and actuation system. 6G will provide a significant contribution by relying on 3D communication platforms that can bring distributed
edge cloud functionalities, e.g. distributed decision mechanisms, {\it on demand}, {\it when and where needed}. In some cases, like autonomous driving, reliable safety mechanisms are essential to prevent accidents. This will require very demanding levels of communication reliability (i.e., above 99.9999) and low end-to-end latency (below 1 ms).  
Moreover, inter-communication among cars will be a key action to reduce risks of accidents. This will require high data rates links between vehicles and between vehicles and road side units. }

\textcolor{black}{
A sustainable development of course pays close attention to energy consumption. Hence, 6G will have to develop truly effective energy-efficient communication strategies. The vision is to achieve, wherever possible, {\it battery-free communications}, targeting communication efficiency in the order of 1 pJ/bit \cite{Memon2019}.
}

\textcolor{black}{The KPI's associated to the previous services are summarized in Table \ref{Table1}, to highlight the necessary improvement with respect to 5G KPI's. Some like delay jitter and energy/bit do not really represent a focus of 5G and then they are not-specified (NS) in 5G, whereas they represent KPI's in 6G.
\textcolor{black}{
\begin{table}[ht]
\centering
\begin{tabular}{ | l | c | c | } 
\hline
KPI& \textbf{5G} & \textbf{6G} \\ 
\hline \hline
Traffic Capacity & 10 Mbps/$m^2$ &  $\sim$ 1-10 Gbps/$m^3$ \\ 
\hline Data rate DL & 20 Gbps &  1 Tbps \\ 
\hline
Data rate UL & 10 Gbps &  1 Tbps \\ 
\hline
Uniform user experience & 50 Mbps 2D everywhere &  10 Gbps 3D everywhere\\ 
\hline
Latency (radio interface) &  1 msec &  0.1 msec \\ 
\hline
Jitter & NS &  1 $\mu$sec \\
\hline
Reliability (frame error rate) & $10^{-5}$ & $10^{-9}$\\
\hline
Energy/bit & NS &  1 pJ/bit\\
\hline
Localization precision & 10 cm on 2D &  1 cm on 3D\\
\hline
\end{tabular}
\caption{Comparison of 5G and 6G KPI's; NS= Not Specified.}
\label{Table1}
\end{table}
}
}

\subsection{Enablers}
Fulfilling the challenging requirements of the new services, 6G will build on a set of technological enablers. We envision the following major thrusts:\\

\textcolor{black}{
{\bf A new architecture: } The need to support nearly deterministic services, as in high-precision manufacturing, guaranteeing very tight physical constraints, like latency or energy consumption, requires a new Internet architecture that combines different resources, such as communication and computation resources, within a single framework. The architecture will include: a {\it new data plane}, able to adapt, dynamically, to different operating modes and to enable holographic communications; {\it a new control plane}, enabling for example the synchronization of concurrent streams for holographic communications, and preferred path routing protocols to establish  nearly deterministic (i.e., with very low jitter) links, to enable high precision manufacturing; {\it a new management plane}, incorporating self-configuration and self-optimization capabilities, leveraging on the strong support of Artificial Intelligence (AI).}\\ 
\textcolor{black}{
{\bf A pervasive introduction of artificial intelligence at the edge of the network: }  Distributed AI algorithms, possibly running under sever delay constraints, are expected to play a key role in various aspects: {\it self-optimization} of network resource allocation, possibly adopting {\it proactive strategies} based on {\it network learning and prediction}; development of smart mobile applications, running either directly on the mobile devices or remotely through computation offloading mechanisms, that learn from user behavior and act as a {\it context-aware virtual intelligent assistant}; development of {\it semantic inference} algorithms and {\it semantic communication} strategies to incorporate knowledge representation in communication strategies. This will be particularly useful for an effective deployment of holographic communications. The role of AI in 6G will be further expanded in Section III;}\\
\textcolor{black}{
{\bf 3D coverage: } The design of a 3D communication infrastructure incorporating terrestrial and aerial radio access points {\it and} mobile edge hosts makes possible to  {\it bring cloud functionalities on demand}, where and when needed. As opposed to the common approach relying on a fixed infrastructure, this strategy is much more economically efficient when the requests are highly varying across space and time, as it happens in case of sporadic events gathering lots of people or in remote areas, in case of disasters. The idea is to manage a plethora of aerial platforms, including Unmanned Aerial Vehicles (UAV), High Altitude Platform Station (HAPS) flying at around 20 Km of altitude,  and constellations of very low Earth orbit (LEO) satellites, flying at altitudes in order of a few hundred Kilometers, in order to bring cloud functionalities under controllable delay constraints.}\\
\textcolor{black}{
{\bf New physical layer incorporating sub-THz and VLC: } The need to support very high data rates, up to Tbps, to enable for example holographic communications, requires the exploitation of  sub-THz bands and visible light communications. This aspect will be further explored in Section IV. {\it Energy-efficient communication} strategies are also expected to become more and more important, especially in view of a pervasive deployment of the Internet-of-Things, with myriads of tiny sensors. Energy harvesting mechanisms and advanced wireless charging technologies and their fundamental limits are presented in \cite{WirelessCharging},  with a focus on promising distributed laser charging techniques showing that wireless charging might approximately deliver 2W of power up to a distance of about 10 meters. An even more drastic approach will rely on the exploitation of {\it ambient backscattering}, which enables tiny devices to operate with no battery, by redirecting ambient radio frequency (RF) signals  (for example using on/off encoding) without requiring active RF transmission \cite{VanHuynh18}.}\\ 
\textcolor{black}{
{\bf Distributed security mechanisms: } The vision delineated so far foresees a huge exchange of data to enable a pervasive use of AI techniques. Clearly, this poses some major challenges in terms of security, privacy, and trust, which need to be properly addressed by 6G networks. 
Innovative cryptographic techniques should be used to get an effective merge of AI and privacy. For example, a mobile user wishing to run a remote machine learning algorithm on its own data, could use {\it homomorphic encryption} \cite{gentry}. In this way, rather than sending the raw data, it could send encrypted data, let the remote machine learning algorithm run on the encrypted data, and still be able to recover the desired output. Existing schemes are not practical yet due to high computation complexity, but we might expect that, in a decade or so from now, the complexity issue could be alleviated. Decentralized authentication is another key issue, especially for the IoT scenario. Distributed Ledger Technologies, exploiting blockchain-like mechanisms, are expected to play a key role for distributed authentication \cite{Ferraro}.}

\section{Pervasive AI and holistic management of C4 resources}

Artificial intelligence (AI), and more specifically machine learning (ML), is already permeating many applications running on today smartphones. It is not difficult to predict an ever increasing usage of learning mechanisms at both network and mobile terminal level. ML cam enable proactive network resource allocation and thus improve performance, especially in delay-sensitive applications. The exploitation of machine learning algorithms at the edge of the network has been recently suggested in  \cite{Park2018}. 
We believe AI will play a key role in 6G in at least the following aspects.\\
\textcolor{black}{
{\bf Semantic communications:} Shannon's classical information theory (CIT) represents the basic foundations of all modern communications. Yet, Shannon deliberately considered semantic aspects as irrelevant. However, now that communications have reached all the limits predicted by Shannon's theory, a true leap forward can come from incorporating semantic aspects in communications \cite{Bao2011}. 
Let us consider as a motivating example a holographic communication, where a speaker gives a speech in a remote place by sending a set of multi-camera video data enabling the reconstruction of her hologram. This requires the transmission of a huge amount of video data. Suppose now that part of this data gets lost, maybe because of blockage events due to the sudden appearance of an obstacle in the millimeter wave, or even worse sub-THz, link. Conventional systems will need to retransmit the lost packets. Conversely, assume that transmitter and receiver share some common knowledge about the speaker. This shared knowledge enables the receiver to reconstruct the hologram, even in the presence of several missing packets. In other words, content information (semantic) can be recovered through semantic inference, if transmitter and receiver share some common knowledge. Similar examples can be made about voice. Recovering the word ``cat'', when the word ``car'' was transmitted, looks like a small mistake at syntactic level, but it represents a big mistake at semantic level. Conversely, recovering ``automobile'' instead of ``car'' is a big mistake at syntactic level, but it is not really a big mistake at semantic level. In summary, semantic communication can greatly improve communication efficiency by assigning different priorities to different data, {\it based on meaning}, and exploiting any form of shared knowledge to enable semantic inference.}\\
\textcolor{black}{
{\bf Machine learning and deep neural networks: }ML will play a key role at the network level by enabling self-organization strategies, including self-learning, self-configuration, self-healing. 
At the terminal level, ML will lead to the deployment of mobile applications that provide a virtual intelligent assistance and learn from the user's behavior. Furthermore, deep neural networks (DNN), with their ability of being a universal function approximator, will make possible to parameterize many network functions with a limited set of parameters (the DNN's coefficients, at different layers, after proper training). This universal function encoder will facilitate the deployment of network functionalities throughout the network by using a general purpose architecture, whose parameters can be periodically updated by running training algorithms offline.}\\
{\bf Holistic management of C4 resources: } A key feature of 6G networks is that its architecture should be designed to handle communication, computation, caching and control (C4) resources as parts of a single system, whose efficient management requires a {\it joint} optimization.
\begin{figure}[h]
\centering
\includegraphics[width=\columnwidth]{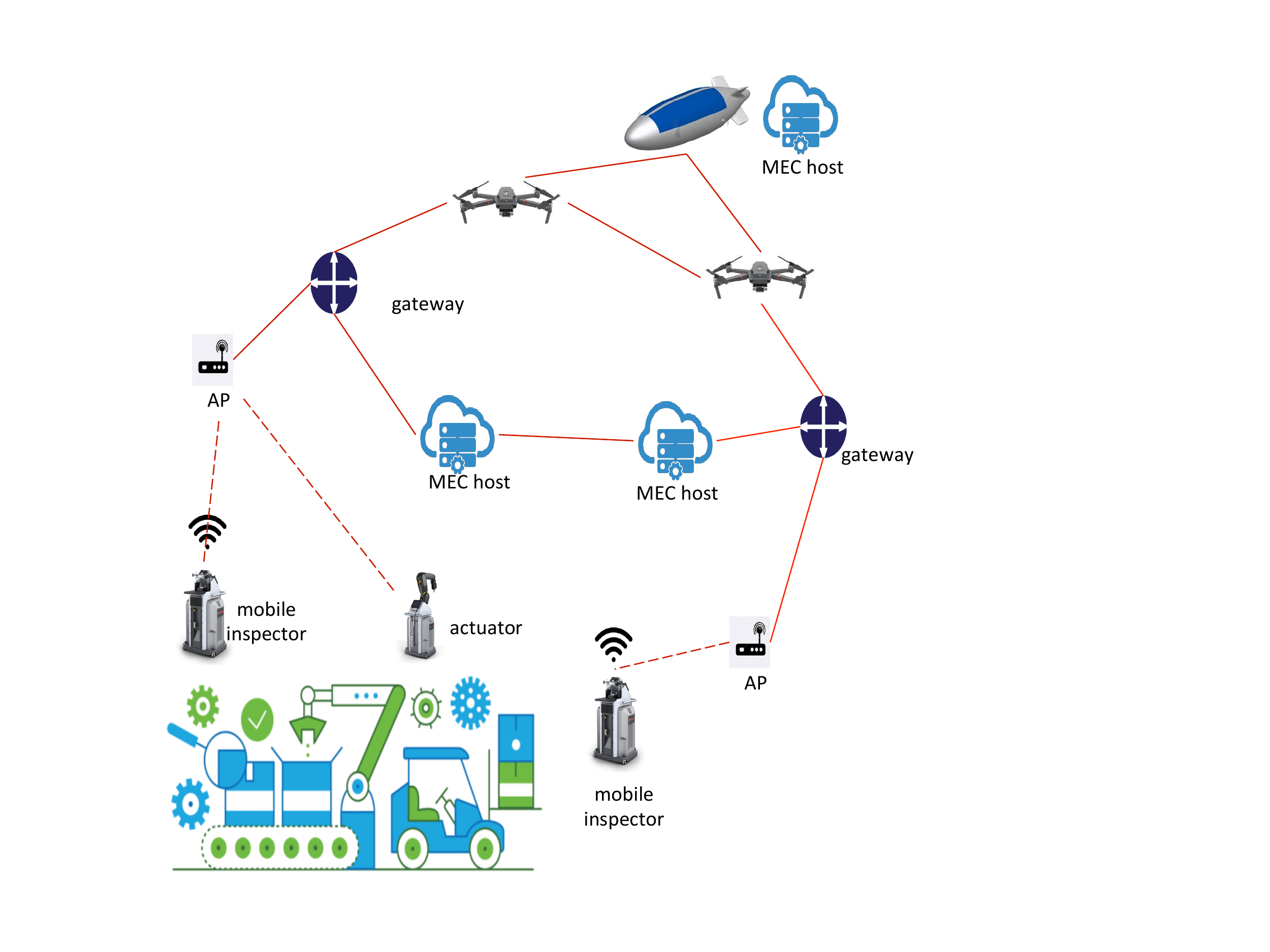}
\caption{C4 scenario incorporating 3D coverage.}
\label{fig:Automated_factory}
\end{figure}
A motivating example is represented, for instance, by the autonomous control of an industrial process in an automated factory sketched in figure \ref{fig:Automated_factory}. In this scenario,  mobile robots take videos of the industrial process and send the data to the edge cloud, where learning algorithms run to detect anomalies and take decisions that are sent back to an actuator aimed to implement the proper countermeasures. The edge cloud may be present in a fixed infrastructure, or it may be made available through aerial platforms. A key performance indicator in this scenario is the end-to-end (E2E) delay, which includes: communication time, i.e the overall time spent to share data among the various actors (robot, cloud, data-base, actuator, etc.); computation time (the time necessary to run the decision algorithm). To 
approach a nearly deterministic control, it is necessary to bring resources, i.e. access points, cloud resources and contents, as close as possible to the end user. This strategy is promoted by the deployment of mobile edge computing (MEC), which is already an important component of 5G. However, in 5G communication and computation resources are handled separately. Conversely, we believe that the inevitable limitation of resources that can be available at the edge of the network calls for a holistic management of all C4 resources. In this new scenario, the assignment of users to access points, dynamic caching and migration of virtual machines will all need to be orchestrated jointly to guarantee the E2E requirements and an efficient use of resources.
The need for a joint C3 (computation, communication and caching) optimization was advocated in \cite{Ndikumana18}, where control was still referring to computation offloading. However the true challenge will be how to enable an intelligent control with stringent E2E delay guarantees, as required in, e.g., autonomous driving, autonomous factory, etc.

\section{Communicating through Sub-TeraHertz and Visible Light Links}
\label{sec:THz-VLC}
The never ending quest for higher data rates, more spectrum and higher network densification sets the target for the future decade of wireless services to provide at least the challenging Tbps aggregated bit rate in small regions. This would represent about two orders of magnitude bit rate increase  with respect to 5G \cite{Netword2020-2015}. 
To tackle this challenging task, 6G will have to exploit different enabling technologies. 
The possible strategies to increase the volume data rate (bits/sec/m$^3$) are: increase bandwidth, by using sub-THz and visible light spectrum; increase spectral efficiency, using advanced MIMO technologies; combine multi-RAT and 3D multi-link connectivity at a scale going well beyond 5G; perform a (ultra) dense deployment of radio access points in a 3D domain, exploiting aerial platforms. 

The European Telecommunications Standards Institute (ETSI) and other agencies are currently considering the bands beyond 90 GHz to provide  high-data rate wireless services for short (5 to 20 meters) and medium distance (50 to 200 meters) fixed links. However, beyond 5G and 6G current research is already focusing on how to efficiently exploit the sub-THz (90- 450 GHz) and visible light bands as solutions to provide optical-fiber like performance from 100 Gbps up to a few Tbps. 



\subsection{The sub-TeraHertz Opportunity for 6G}
 \label{sec:THz}

 Sub-THz communications are envisioned for beyond 5G communications at both device access and network backhauling and frounthauling. 
For example, we expect that, for device access communications with (ultra) dense deployments of access points within 10 meters of line of sight communication range, one Tbps is achievable in principle by a single link  with 250 GHz of carrier frequency (assuming 20\% fractional bandwidth (Bw), 40 dBi antenna gain at both transmitter and receiver and, and a noise figure of 20 dB) but it requires very high spectral efficient modulation (and coding) schemes. While in future, the evolution of transceiver and antennas technologies will provide solutions for achieving such challenging performance, the associated energy consumption still remains a severe limitation for the applicability of such technologies for battery-powered terminals. 
 In a first generation of sub-THz communications,  most likely, simple modulations schemes with low spectral efficiency will be used to communicate at very short distances by using large amounts of BW with very simple transceiver architectures, therefore avoiding power-hungry complex digital signal processing and coding. Then, a technological breakthrough will be needed otherwise the peak data rate will be limited by the available supply power.
 
\begin{figure}[h!]
\begin{centering}
	\includegraphics[width=\columnwidth]{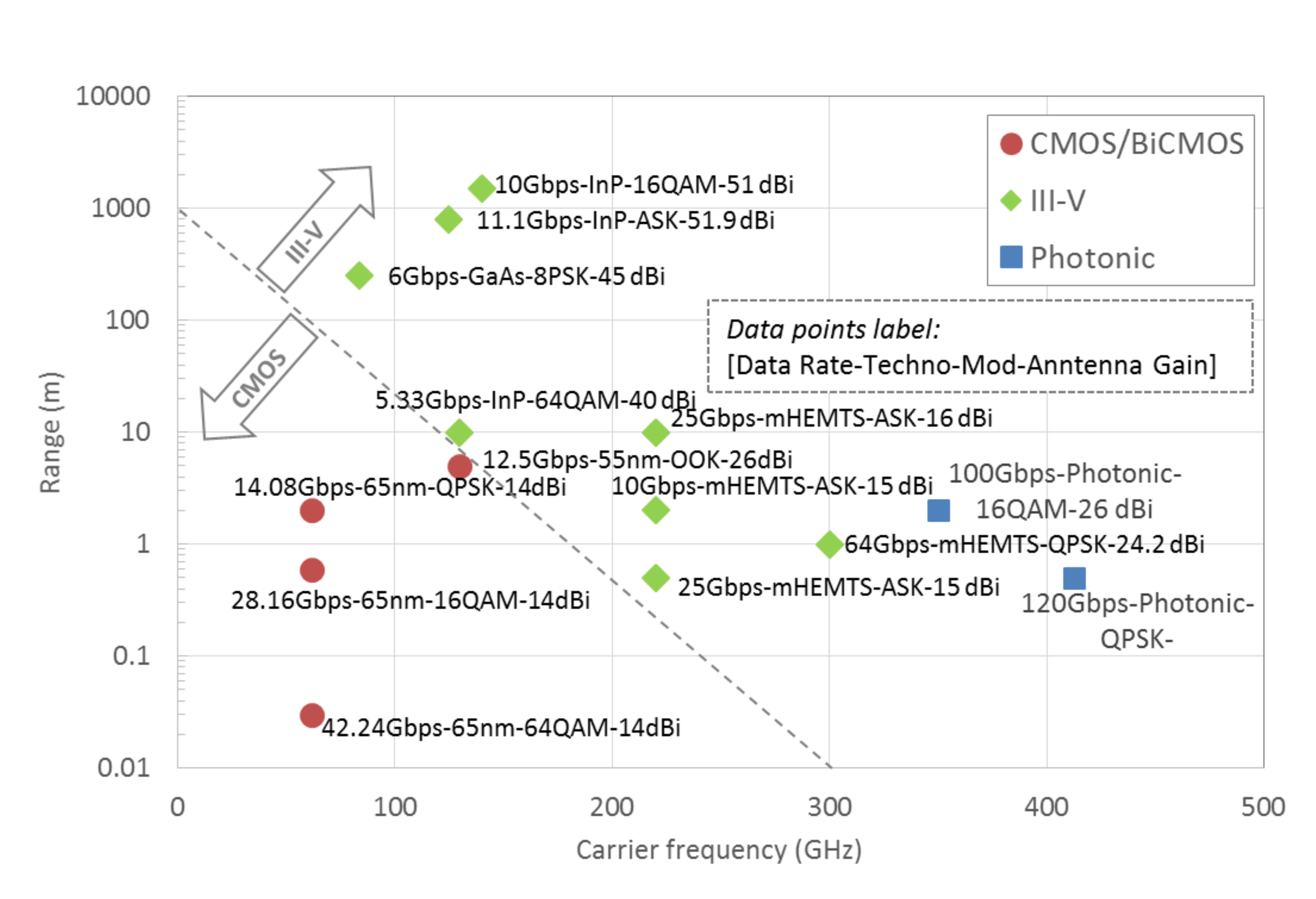}
\caption{Sub-THz Hardware IC Tecnologies.}
\label{fig:THz-HW-Tecnologies}
\end{centering}	 
\end{figure}

The first exploitation of sub-THz communications  for fronthauling and backhauling links will require, even with ultra densification of 5G and 6G networks, to cover at least between tens to hundreds of meters. Nevertheless, for these challenging distances, at such high frequencies, the single link peak capacity falls well below 1 Tbps. After all, the Graal of the Tbps can be reached also at such communication ranges allowing more complex and less power constrained communications, engineering antenna array with higher gains and investing in more hardware complexity for combining multiple parallel links (of the order of 100s of Gbps) using diversity (polarization, frequency, spatial, ...). 
 Achieving several hundred Gbps over a single wireless links is a today research and prototype reality.
 Figure \ref{fig:THz-HW-Tecnologies} shows recent realizations of multi-Gbps wireless transceivers implemented using different integrated circuit (IC) technologies. The points on the figure include labels displaying the data-rate, the technology, the modulation scheme and the antenna gain (if available). The trade-off between carrier frequency and range is clearly shown in the figure. The larger carrier frequency solutions use a large amount of RF bandwidth in a single RF channel and simple modulations schemes such as Amplitude-Shift Keying (ASK)/ Binary Phase SK (BPSK) which allows for digital-less demodulation. Such transceivers have been demonstrated for carrier frequencies up to 300 GHz and achieve communication in the range of 1 m up to 1 km using high gain antennas, as shown on Figure \ref{fig:THz-HW-Tecnologies} \cite{Jimenez}. 
On the other hand, CMOS based transceivers are limited to lower frequencies and bandwidths. They use higher order modulations such as Quadrature PSK (QPSK), 16QAM or even 64QAM. This second type of transceivers have demonstrated high data-rates at lower carrier frequencies, comparable to those of much larger bandwidth III-V transceivers, as shown on figure \ref{fig:THz-HW-Tecnologies}.  
Recent photonic-based modulation transceivers  have also been included in the figure, and correspond to the highest carrier frequencies, but their degree of integration if currently very low. Today's advanced digital baseband systems data-rate processing capabilities are somehow limited to a few tens of Gbps, and are also power consumption bonded. There is indeed a trade-off to explore between complexity/power consumption of transceiver and antennas and the exploitation of diversity techniques (i.e. number of single links) to reach the (aggregated) Tbps target. For instance, a reasonable trade-off needs to be done between single link BW and the overall number of channels required to cover the full RF band to provide the desired throughput. With the advances of materials and technology such trade-off might be relaxed opening cost effective Tbps communications in 6G networks.


\subsection{The Visible Light Communications Opportunity for 6G}
 \label{sec:VLC} 
\begin{figure}[h!]
\begin{centering}
\includegraphics[width=\columnwidth]{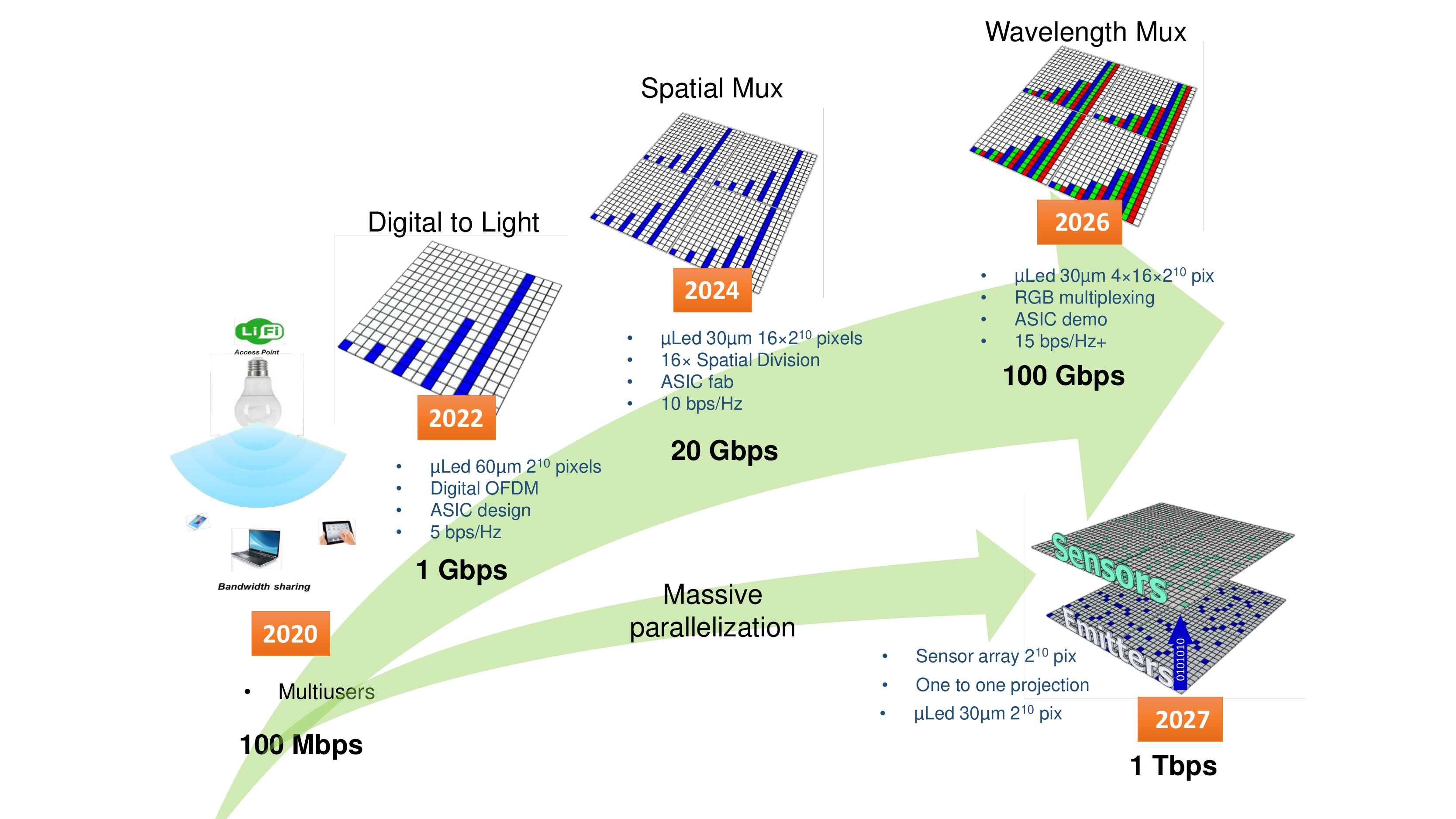}
\caption{Visible Light Communication Roadmap from Mbps to Tbps.}
\label{fig:VLCRoadMap}
\end{centering}	 
\end{figure}

In order to provide optical-fiber like performance, one possible complementary technology for future 6G networks is to exploit the visible spectrum with visible light  communication (VLC) techniques for short range (up to few meters) links,  which compared to classically adopted RF bands offers ultra-high bandwidth (THz), zero electromagnetic interference, free unlicensed abundant spectrum, very high frequency reuse \cite{Haas-VLC2018}. Today, currently available VLC indoor technology is clamped from several tenths of Mb/s to 100 Mb/s in the range of about 5 meters. Such limitation - due to commercial over the shelf (COTS) light fixtures and absence of beamforming - is forecast to disappear with the introduction of upcoming new light sources based on microLED. Such innovative components will soon offer 1+GHz bandwidth and near 10Gb/s have been achieved in lab on a single diode LED. In our vision, VLC has a huge potential of improvement in the next decade, catching up the current  performance gap between VLC and 5G technologies that already proved the Gbps experience \cite{5GCHAMPION2018} and reaching the Tbps with VLC technologies when 6G services will be launched. In our vison, VLC will be able to provide short range indoor connectivity reaching the Tbps with full operational available demonstrators maturity by 2027 (see figure \ref{fig:VLCRoadMap}). The first milestone will be in 2019, by the launch of 5G services, when full 1 Gbps indoors short range capability will be demonstrated with thousands of LED active sources for lighting functionality. Then, a new challenge will appear at short term in order to pilot such huge matrices and preserve the high bandwidth. To this end, it will be required the hybridization of microLED matrices and CMOS driver arrays into a single chip. Then, in order to reach by 2024 the today claimed highest 5G link capacity (20 Gbps), it will be required a more complex device with the parallelization of several aforementioned chips and the introduction of a dedicated imaging optical system leading to a spatial separation of users, also known as optical beamforming, with a significant increase of cellular throughput. Eventually, by 2026 when both microLED technologies and spatial multiplexing techniques will be mature and cost effective, white light based on different wavelengths will unlock throughput thanks to wavelength division multiplexing leading to potentially 100+ Gb/s for ultra-high data rate VLC access points. In 2027, adding on top of the above mentioned technologies massive parallelization of microLED arrays, the target Tbps aggregated throughput will be available.

\section{Conclusions}
\textcolor{black}{
In this paper, we highlighted some of the new services and their technological enablers that should represent the core of next generation (6G) networks. At the physical layer, we believe that the level of maturity reachable in a decade from now by sub-THz and visible light communication can make these technologies a powerful enabler. A more massive use of multi-RAT and multi-link techniques will be needed, to tackle the challenges arising in the propagation in these higher frequency bands and providing high reliability links. At a higher level, we think that new architectures will need to be designed to really bring delay-sensitive AI applications to the end user, pushing for a major integration and distributed optimization of communication, computation, caching and control resources. We mentioned possible ways to go beyond Shannon's classical information theory by incorporating semantic aspects. Finally, looking at another law that has been able to predict the miniaturization capabilities of integrated circuits (IC) for a long time, i.e. Moore's law, we predict that at some point, Moore's law will inevitably break, but this will not end the increase of computational capabilities of electronic devices, as most sophisticated device interactions between IC's, at different scales, will more than complement the limitations of the single IC.
}
\begin{IEEEbiography}{Emilio Calvanese Strinati}
is the Smart Devices \& Telecommunications Scientific and Innovation Director at CEA-Leti / Minatec Campus. In 2001 he obtained his telecommunication engineering master degree from the University of Rome ‘La Sapienza’, his Ph.D in Engineering Science in 2005 from Telecom Paris Tech and since 2018 he holds the French Research Director Habilitation (HDR). He has published around 100 papers in international conferences, journals and books chapters, given more than 100 international invited talks and keynotes and is the inventor of more than 60 patents.
\end{IEEEbiography}
\begin{IEEEbiography}{Sergio Barbarossa}
Sergio Barbarossa, received his MS and Ph.D. EE degree from the Sapienza University of Rome, where he is now a Full Professor. He is an IEEE Fellow and EURASIP Fellow. He has been an IEEE Distinguished Lecturer in 2013-2014. He received the 2000 and 2014 IEEE Best Paper Awards from the IEEE Signal Processing Society. In 2010 he received the Technical Achievements Award from the European Association for Signal Processing (EURASIP). Since 2000, he has been the scientific coordinator of several EU projects on wireless sensor networks, small cell networks, and distributed mobile cloud computing. He is currently the technical manager of the H2020 Europe/Japan project on "Millimeter-wave edge cloud as an Enabler for 5G" (5G-MiEdge), merging millimeter wave and mobile edge computing technologies. His current research interests are in the area of millimeter wave communications and mobile edge computing, graph signal processing, machine learning and distributed optimization.
\end{IEEEbiography}

\begin{IEEEbiography}{Jose Luis González-Jiménez} 
(M’99–SM’12) received the Diploma in telecommunication engineering from Ramon Llull University, Barcelona, Spain, in 1992, and both the M.S. degree in telecommunications engineering and the Ph.D. degree in electronic engineering from the Universitat Politecnica de Catalunya (UPC), Barcelona, in 1994 and 1998, respectively. He received the “Habilitation à Diriger des Recherches” degree from the Grenoble University, France, in 2013.
He is currently a Senior Expert and RFIC and mmW Research Engineer at CEA/LETI, Grenoble, France, and invited lecturer at Phelma Engineering School, Grenoble-Alpes University. He is the author of two books, a book chapter, 32 international journal papers, and more than 80 conference papers. He holds 15 patents. His research interests include very large-scale integration design, mixed-signal/RF and mmW ICs, silicon photonics, and signal and power integrity in SoC and RFICs
\end{IEEEbiography}

\begin{IEEEbiography}{Dimitri Kténas}
is head of Wireless Technology Department at CEA-Leti / Minatec Campus since 2018, focusing on algorithm and HW/SW implementation of digital signal processing and protocols for 5G and antenna design. He has co-authored 60+ papers in international journals and conference proceedings and 5 book chapters, and is the main inventor or co-inventor of 13 patents related to 5G. He has been advising students for 10 years, including PhD students and post-doctoral research fellows. 
\end{IEEEbiography}
\begin{IEEEbiography}{Nicolas Cassiau}
received the MS degree in Signal Processing in 2001 from Polytech’Nantes, France. Since then he is a Research Engineer and project manager at CEA-Leti in Grenoble, France. His fields of interest are digital wireless communications and algorithms design. He has been working in particular on physical layer design and assessment for 4G and 5G. He has authored or co-authored over 30 papers and holds several patents in the above-mentioned fields.
\end{IEEEbiography}
\begin{IEEEbiography}{Luc Maret}
is a permanent research scientist and project manager at CEA-Leti / Minatec Campus since 2002 within the High-Throughput Systems Laboratory. He has been involved in physical layers designs and algorithms for wireless systems mainly in millimeter wave. During the last 6 years, he has focused on PHY layers design and implementation of optical wireless communications and has conducted VLC/LiFi activities at CEA/Leti. He has been advising students for about 10 years.
\end{IEEEbiography}

\begin{IEEEbiography}{Cedric Dehos}
joined CEA Leti in 2003 after graduation from ESIEE Paris in Telecommunication and Signal Processing. Since then he has been involved in system level design of complex CMOS RF and digital base band circuits. He has been involved in various mmw developments, including Short Range Radar, and WiGig. In 2011 he led Leti mmw designs and began moving the developments towards 5G small cells and short range chip to chip communications.
\end{IEEEbiography}
\end{document}